\documentstyle[12pt]{article}
\newcommand{\too}{\rightarrow}
\newcommand{\EQ}{\begin{equation}}
\newcommand{\EN}{\end{equation}}
\newcommand{\EQA}{\begin{eqnarray}}
\newcommand{\EQN}{\end{eqnarray}}
\newcommand{\beq}{\begin{equation}}
\newcommand{\eeq}{\end{equation}}
\newcommand{\bea}{\begin{eqnarray}}
\newcommand{\eea}{\end{eqnarray}}
\newcommand{\bean}{\begin{eqnarray*}}
\newcommand{\eean}{\end{eqnarray*}}
\newcommand{\dz}{\delta^D (0)}

\newcommand{\E}{{\rm e}}
\renewcommand{\thefootnote}{\dag}

\newcommand{\B}{\beta}

\newcommand{\D}{\Delta}
\newcommand{\k}{\kappa}

\renewcommand{\l}{\lambda}

\newcommand{\lt}{\left(}
\newcommand{\rt}{\right)}

\newcommand{\ta}{\tilde{a}}
\newcommand{\TA}{\tilde{a}}
\newcommand{\tx}{\tilde{x}}
\newcommand{\tk}{\tilde{\kappa}}
\newcommand{\TX}{\tilde{X}}
\newcommand{\vp}{\varphi}
\newcommand{\dx}{ \frac{d^D X}{(2\pi)^{D/2}} }
\newcommand{\dxx}{ \frac{d^D X'}{(2\pi)^{D/2}} }
\newcommand{\tidx}{ \frac{d^D \tilde{X}}{(2\pi)^{D/2}} }
\newcommand{\TPA}{\tilde{\partial}}

\newcommand{\PA}{\partial}
\newcommand{\NB}{\frac{N}{\B}}

\newcommand{\HF}{\frac{1}{2}}
\newcommand{\QR}{\frac{1}{4}}

\newcommand{\lb}{\mbox{[}}
\newcommand{\rb}{\mbox{]}}
\newcommand{\BL}{\mbox{[}}
\newcommand{\BR}{\mbox{]}}

\newcommand{\la}{\langle}
\newcommand{\ra}{\rangle}
\newcommand{\PH}{{\bf \phi}^2}

\begin{document}
\begin{titlepage}
\begin{flushright}
TIT/HEP-229 \\
hepth@xxx/9310193\\
August 1993
\end{flushright}

\begin{center}

\ \

\ \

\Large

{\bf Scaling Violation in O(N) Vector Models}\\

\vspace{2.5cm}

\large

{\sc	Shinsuke Nishigaki}

\vspace{2.0cm}
\renewcommand{\thefootnote}{\ }
{\it
Department of Physics, Tokyo Institute of Technology\\
O-okayama, Meguro, Tokyo 152, Japan }\footnote{
e-mail address: {\sf nsgk@phys.titech.ac.jp}}
\setcounter{page}{0}

\vspace{4cm}
\large
{\bf Abstract}\\
\end{center}
We investigate $O(N)$-symmetric vector field theories
in the double scaling limit. Our model describes branched polymeric systems
in $D$ dimensions, whose multicritical series interpolates
between the Cayley tree and the ordinary random walk.
We give explicit forms of residual divergences in the free energy,
analogous to those observed in the strings in one dimension.
\normalsize
\end{titlepage}

\baselineskip=0.6cm
\begin{center}
{\large {\bf  1. introduction}}\\
\end{center}

The method of $1/N$ expansion had provided us a insight into non-perturbative
aspects of field theories, especially of gauge field theories \cite{ZJ1}.
Its utility, however, as a constructive definition of the string theory
in less than one dimension is too great to be called as a by-product;
matrix models in the double scaling limit \cite{DSL} turned out to be equipped
with the richest structure, and to share the physical content of the
continuum theory \cite{GiMo}.

In ref.\ \cite{NY1}, we have presented another example of non-perturbative
treatment of random statistical objects ---
branched polymers via double-scaled $O(N)$-symmetric vector models.
Our interest in these models was twofold: as a toy model for
investigating non-perturbative features of bosonic strings in the
branched-polymeric phase, and more practically, as a probe into the structure
of two-dimensional gravity through the strong resemblance between the two
theories. In particular, the fact that the Virasoro constraints on
the $\tau$-function is characterized as a symmetry of the `string equation'
\cite{Y2} is readily observed in its vector-model
counterpart, owing to the simplicity and linearity of the latter \cite{NY2}.

The notion of double scaling limit renewed the approach to the $O(N)$-symmetric
field theories which has been a fundamental set-up for the $1/N$ expansion.
In refs.\ \cite{ZJ2} a local field theory of the composite field was
extracted via the double scaling limit, and the `canonical' combination
of scaling parameters was shown to be violated through the renormalization
of ultraviolet divergences.  See ref.\cite{Sch} for the supersymmetric
generalization. It is also pointed out \cite{Y3} that the form of double-scaled
field theory is universal, scarcely dependent upon the
original $O(N)$ field theory, i.\ e.\ how we discretize and weight
the branched polymers.

The purpose of this paper is to clarify
the violation of the canonical scaling, in particular residual divergences
in the free energy which is not explicitly worked out in refs.\cite{ZJ2}.
For completeness we review the double scaling limit of
$O(N)$-vector field theories.  Critical exponents of
the theory, which fulfill Fisher's relation, suggest that the statistical
systems at hand are thought of multicritical generalizations of the
Cayley tree.
Then we state our main result on the scaling violations of
the renormalized free energy at low-loop levels.\\
\begin{center}
{\large {\bf  2. branched polymers and O(N) vector models}}\\
\end{center}
\setcounter{equation}{0}

We start by
considering the one-dimensional reduction of Polyakov string:
\beq
F=\sum^{\infty}_{g=0} \kappa^{2g-2} \
\sum_{(BP_{g})} \E^{-t \cdot \mid BP \mid}
\int \lb d^D X(\tau) \rb \  \E^{-\HF \int  d \tau (dX^\mu / d\tau)^2}.
\eeq
\label{2.2}
Here the sum $(BP_g)$ is taken over the set of connected branched polymers
with $g$-loops, $ \vert BP \vert$ denotes the total {\it intrinsic} length of a
branched polymer,
$t$ the cosmological constant and $\tau$ parametrizes the polymer.
In our context an arbitrary
polymer should be weighted solely by its length and number
of loops, irrespective
of its way of branching.

The above definition is only formal unless accompanied
with a definite choice of
measure as well as of regularization. We employ random-lattice regularization
of branched polymers, where polymers are made of line segments (bonds)
with the same lattice
constant $\tilde{a}$ by joining {\it arbitrary} numbers of the ends at their
vertices (molecules).
Matter degrees of freedom in eq.(1) are assigned on the vertices of
the polymer. Then the regularized partition function is defined as
\bea
F&=&
\begin{picture}(20,30)(-10,-3)
\setlength{\unitlength}{0.06cm}
\put(0,0){\circle{2}}
\put(10,0){\circle{2}}
\put(1,0){\line(1,0){8}}
\end{picture}
\ \ \ \ \ + \
\begin{picture}(20,40)(-10,-3)
\setlength{\unitlength}{0.06cm}
\put(0,0){\circle{2}}
\put(10,0){\circle{2}}\put(1,0){\line(1,0){8}}
\put(20,0){\circle{2}}\put(11,0){\line(1,0){8}}
\end{picture}
\ \ \ \ \ \ \ \ \ + \cdots + \
\begin{picture}(40,40)(-10,-3)
\setlength{\unitlength}{0.06cm}
\put(1,0){\line(1,0){8}}\put(0,0){\circle{2}}
\put(11,0){\line(1,0){8}}\put(10,0){\circle{2}}
\put(21,0){\line(1,0){8}}\put(20,0){\circle{2}}
\put(31,0){\line(1,0){8}}\put(30,0){\circle{2}}
\put(41,0){\line(1,0){8}}\put(40,0){\circle{2}}
\put(51,0){\line(1,0){8}}\put(50,0){\circle{2}}
\put(20,1){\line(0,1){8}}\put(60,0){\circle{2}}
\put(20,11){\line(0,1){8}}\put(20,10){\circle{2}}
\put(21,20){\line(1,0){8}}\put(20,20){\circle{2}}
\put(31,20){\line(1,0){8}}\put(30,20){\circle{2}}
\put(40,11){\line(0,1){8}}\put(40,20){\circle{2}}
\put(40,01){\line(0,1){8}}\put(40,10){\circle{2}}
\end{picture}
\ \ \ \ \
\ \ \ \ \
\ \ \ \ \
\ \ \ \ \
\ \ + \cdots
\nonumber \\
& = &
\sum^{\infty}_{g=0} \kappa^{2g-2}_0 \
\sum_{(BP_{g})} \frac{1}{S(BP)}\ \E^{-t_0 \cdot n_1 (BP) }
\int \prod_{i\in BP} d^D X_i \ \E^{-\HF \sum_{\la ij \ra} \mid X_i - X_j
\mid^2}
\label{2.3}
\eea
where $S(BP)$ denotes the symmetry factor of a branched polymer,
$n_1 (BP)$ the number of bonds in a polymer and
$\sum_{\la ij \ra}$ the sum over adjacent vertices, and the subscript $0$
refers to bare quantities.
A possible generalization of this model is to introduce branch-units with more
than three end points, with activities $\lambda_n$.
This generalization is imported from the matrix-model realization of
two-dimensional gravity where inclusion of equilateral polygons
into the dynamical triangulation generates multi-critical behaviours.

These statistical systems of branched
polymers are exactly realized as dual diagrams
of Feynman graphs in the perturbative expansion of the following
$O(N)$-symmetric vector model \cite{Y3} ,
\beq
Z (\lambda) =\int \left[  d^N {\bf \phi} \right] \ \exp -\B \lt
\int d^D X \ V( \PH (X) )
-\frac{\l}{4}\int d^D X d^D X' \ \E^{-\HF
 \mid X-X' \mid^2} \PH (X) \PH (X') \rt ,
\label{2.6}
\eeq
where
$\phi=(\phi_a)$ is $N$-component real scalar field and
\beq
V(\PH ) = \HF \PH -\sum_{n\geq 2} \frac{\lambda_n}{2^n} (\PH )^n .
\eeq
Here a dual diagram is meant in a sense of one-dimensional geometry, i.\ e.\
the 0-simplex ($({\bf \phi}^2)^n$ vertex) and
1-simplex (propagator) of Feynman graphs are interchanged
with the 1-simplex (bond or
branch unit) and
0-simplex (molecule) of branched polymers respectively.
Namely, for the simplest potential $V(\phi^2)=\HF \phi^2$, we observe
\bea
\ln \frac{Z (\lambda)}{Z (0)} &=& \
\begin{picture}(20,30)(-10,-3)
\setlength{\unitlength}{0.06cm}
\put(0,0){\circle{10}}\put(5,0){\circle*{1.5}}
 \put(10,0){\circle{10}}
\end{picture}
\ \ \ \ \ + \
\begin{picture}(20,40)(-10,-3)
\setlength{\unitlength}{0.06cm}
\put(0,0){\circle{10}}  \put(5,0){\circle*{1.5}}
 \put(10,0){\circle{10}}\put(15,0){\circle*{1.5}}
 \put(20,0){\circle{10}}
\end{picture}
\ \ \ \ \ \ \ \ \ +
 \cdots + \
\begin{picture}(40,40)(-10,-3)
\setlength{\unitlength}{0.06cm}
\put(0,0){\circle{10}}\put(5,0){\circle*{1.5}}
\put(10,0){\circle{10}}\put(15,0){\circle*{1.5}}
\put(20,0){\circle{10}}\put(25,0){\circle*{1.5}}
\put(30,0){\circle{10}}\put(35,0){\circle*{1.5}}
\put(40,0){\circle{10}}\put(45,0){\circle*{1.5}}
\put(50,0){\circle{10}}\put(55,0){\circle*{1.5}}
\put(60,0){\circle{10}}\put(20,5){\circle*{1.5}}
\put(20,10){\circle{10}}\put(20,15){\circle*{1.5}}
\put(20,20){\circle{10}}\put(25,20){\circle*{1.5}}
\put(30,20){\circle{10}}\put(35,20){\circle*{1.5}}
\put(40,20){\circle{10}}\put(40,15){\circle*{1.5}}
\put(40,10){\circle{10}}\put(40,5){\circle*{1.5}}
\end{picture}
\ \ \ \ \
\ \ \ \ \
\ \ \ \ \
\ \ \ \ \
\ \ + \cdots \nonumber \\
& =&
\frac{1}{S(BP)} \lt N \delta^D (0) \rt^{n_0 (BP)}
\int \prod_{i \in BP} d^D X_i \prod_{\la ij \ra \in BP}
\lt \frac{\l}{\B} \E^{-\HF \mid X_i - X_j \mid^2} \rt  \\
&=&
\frac{1}{S(BP)} \lt N \delta^D (0) \rt^{1-g}
\lt \frac{N\delta^D (0)}{\B} \l \rt^{n_1 (BP)}
\int \prod_{i} d^D X_i
\E^{ -\HF \sum_{\la ij \ra } \mid X_i - X_j \mid^2 } . \nonumber
\eea
Since the divergent factor $\delta^D (0)$ (a redundant $\delta$-function in a
index loop) is always associated with $N$ large, it can be
absorbed into the definition of the bare parameters in eq.(\ref{2.3}).
We then confirm $F=\ln \bigl( Z(\lambda)/Z(0)
\bigr)$ under identification of the parameters
\beq
\kappa_0^2 = \frac{1}{N\delta^D (0)} \ , \
t_0=-\ln \lt \frac{N\delta^D (0)}{\B} \l \rt . \\
\eeq

\begin{center}
{\large {\bf  3. continuum limit}}\\
\end{center}

Now we repeat the method developed in the zero-dimensional case:
First we find a large-$N$ saddle point of the effective potential.
Then we scale the bare cosmological and string coupling constants
towards the critical point simultaneously,
in such a way that each loopwise infrared
divergence of the free energy is compensated by the increasing power of $N$
so that contributions from arbitrarily looped configurations
survive the continuum limit.
We expect a crucial difference to appear between the cases $D=0,1$ and
$D\geq 2$, since in the latter case the canonical scaling determined
by a na\"{\i}ve large-$N$ expansion is altered by the renormalization
procedure of {\it ultraviolet} divergences in the target space.

For brevity we take the simplest model $V(\phi^2)=\HF \phi^2$.
After integrating over angular variables and exponentiating the jacobian
involved, the partition function takes the form
\bea
Z&\!\!\!\!=\!\!\!\! & \int \left[  dx \right] \ \E^{-S_{eff} \left[ x \right]},
 \ \  \ \
S_{eff} \left[ x \right] \equiv \B \left\{
\int \dx \ \HF x (X) \right. \label{eq3.9} \\
- \int \dx \dxx &\!\!\!\! \ \!\!\!\! &\left. \frac{\l}{4}\
\E^{-\HF \mid X-X' \mid^2} x (X) x (X')
-\NB \dz \int \dx \HF \ln \lt \frac{x(X)}{2} \rt
\right\} . \nonumber
\eea
Here we have rescaled $\PH =x$ from eq.(\ref{2.6}) by a factor
$(2\pi)^{-D/2}$ for later convenience.

The large-$N$ saddle point equation of the effective potential reads
\beq
\NB \dz =
x (X) -\l \int \dxx \ \E^{-\HF \mid X-X' \mid^2} x (X) x (X') .
\label{6.1.3}
\eeq
Eq.(\ref{6.1.3})
reduces to the saddle point equation of the one-vector model in $m=2$
criticality, $\D = (1-x/x_c)^2$ ($\D \equiv 1-\NB \dz$)
under the choice $\l=1/4$ and $x_c = 2$ (position-independent saddle point
$x(X)=x_s$ is assumed).
Note that the condition
$\delta S_{eff}/\delta x=
\delta^2 S_{eff}/\delta x^2=0$ which determines the critical values of
$\l$ and $x$ leads to the masslessness of a boundstate
associated with the composite field $x=\PH$.

By substituting the saddle point solution $x(X)=x_s=2(1-\Delta^{1/2})$,
we obtain the leading singular term of the free energy par
unit volume $=(2\pi)^{D/2}$
\beq
F^{(0)} \sim -\frac{1}{3} N \dz \D^{3/2} + {\small \rm regular\ terms}.
\eeq
We can also generate
multicritical behaviours characterized by $\gamma^{(0)}=1-1/m$
by fine-tuning the potential.

The effective action in eq.(\ref{eq3.9}) is expanded in the $1/N$ series
around this saddle point. In terms of $\tx (X)=x(X) -x_s$ it reads
\bea
S_{eff} \left[ \tx \right] &=& \B \int \dx \lt
-\frac{1}{3}\D^{3/2} + \frac{1}{32} (\PA_\mu \tx)^2 + \frac{1}{8}
\D^{1/2} \tx^2 - \frac{1}{48} \tx^3 \rt \nonumber \\
&+& \lt \mbox{\small higher order terms in}\ \tx ,\
\mbox{\small its derivatives and}\ \D \rt .
\label{6.1.4}
\eea
This seemingly local expression follows from the Taylor expansion of
$\tx (X')$ in the nonlocal term in eq.(\ref{eq3.9}),
\[
\int \dxx \ \E^{-\HF \mid X-X' \mid^2} \tx (X') = \tx (X)
+ \HF \PA_\mu^2 \tx (X) + \cdots .
\]
\renewcommand{\thefootnote}{\ddag}
Now we introduce the lattice spacing $a$ in the target space, with which
a dimensionful (physical) length $\TX$ is defined as
\mbox{$\TX = a\cdot X$}.
Then the form of the expression (\ref{6.1.4}) suggest that
the effective action can be made nontrivial and finite
in renormalized quantities (in the sense of continuous polymers)
under the following scaling limit:
\beq
\left\{ \begin{array}{l}
                    \B=\TA^{(D-6)/4}\cdot \kappa^{-2} \\
                    \D=\ta \cdot t \\
                    \tx=\TA^{1/2} \cdot (-4\kappa) \vp
\end{array} \right.
\label{6.1.5}
\eeq
accompanied with the identification between the intrinsic and extrinsic
lattice constants by $\ta =a^4$ approaching zero.
The first equation in (\ref{6.1.5}) requires
the dimension of the space to be smaller than six in order for $\B$
to diverge to infinity so that large-$N$ expansion is meaningful.
Then higher order terms in eq.(\ref{6.1.4})
are suppressed by positive powers of $a$, to render the effective action
to a {\it local}\ field theory
\beq
S_{eff} \left[ \vp \right] = \int \tidx \lt
-\frac{1}{3} \frac{t^{3/2}}{\k^2} + \frac{1}{2} (\TPA_\mu \vp)^2
+ 2t^{1/2} \vp^2 + \frac{4\k}{3}  \vp^3 \rt .
\label{6.1.6}
\eeq

The correlation length of branched polymers
in the target space is characterized by the mass of the $\vp$-field,
$m(t)=2t^{1/4}$. This immediately shows that the (external) Hausdorff dimension
$D_H$,
defined as a diverging rate of the correlation length
\mbox{$\xi (t) \sim t^{-1/D_H}$} in the critical region,
is equal to 4. We observe that these value of critical exponents
($\gamma=\HF$ and $\nu=1/D_H=\QR$) are coincident with those of the Cayley
tree,
as naturally expected from the form of the $\varphi^3$ field
theory derived in the scaling limit.
We point out, moreover, that under the translation $\vp \too \vp - 2t^{1/2}/\k$
the effective action takes the typical form
\beq
S_{eff} \left[ \vp \right] = \int \tidx \lt
\frac{1}{2} (\TPA_\mu \vp)^2 - \frac{t}{\k} \vp + \frac{4\k}{3}  \vp^3 \rt
\label{6.1.7}
\eeq
of a massless scalar field theory perturbed by the most infrared-divergent
operator $\vp^1$, coupled to the cosmological constant;
it reduces to the zero-dimensional `effective action'
in the Laplace-transformed solution to the chain equation in ref.\cite{NY1}
 if the $\vp$-field is translationally invariant.

Now that it is straightforward to extract an $m$-th multicritical
field theory
\beq
S_{eff} \left[ \vp \right] = \int \tidx \lt
\frac{1}{2} (\TPA_\mu \vp)^2 - \frac{t}{\k} \vp + \frac{2^m \k^{m-1}}{m+1}
\vp^{m+1} \rt
\label{6.1.8}
\eeq
out of a vector model with a generalized potential
by fine-tuning $(m-1)$ parameters as in ref.\ \cite{NY1}.
Here the scaling limit
\beq
\left\{ \begin{array}{l}
             \B=\TA^{\lt \lt m-1 \rt D - 2 \lt m+1 \rt \rt / (2m)} \cdot
             \kappa^{-2} \\
             \D=\ta\cdot t \\
             \tx=\ta^{1-1/m} \cdot (-4\kappa) \vp
\end{array} \right.
\label{6.1.9}
\eeq
is accompanied with the identification between the lattice constants
$\ta = a^{2m/(m-1)} \too 0$.
Again the double scaling limit is possible only for $D < 2(m+1)/(m-1)$,
i.\ e.\ exactly when the resulting local field theory is superrenormalizable.

The external Hausdorff dimension is also read from eq.(\ref{6.1.8})
and is equal to
\beq
D_H = \frac{2m}{m-1},
\eeq
which coincides with the results of different approaches
in refs.\cite{HT,ADJ}.
We note that this result seems to suggest that the theory approaches
the ordinary random walk (where $D_H =2$) in the limit $m \too \infty$.
On the other hand, since the Feynman propagator of scalar field theories
behaves as $ \vert \TX \vert^{-(D-2)} $ at the vanishing mass
for large $\vert \TX \vert$,
the anomalous dimension $\eta$ is equal to zero, irrespective of order of
the criticality.
It is instructive to point out that by combining $\gamma^{(0)} = 1-1/m$,
$\nu =(m-1)/2m$ and
$\eta = 0$ we can confirm that Fisher's relation
\beq
\gamma=\nu (2-\eta)
\eeq
between critical exponents is fulfilled for any branched-polymeric systems
of our kinds, as long as these exponents are well-defined. \\

\begin{center}
{\large {\bf 4. renormalization and residual divergences}}\\
\end{center}

We have identified superrenormalizability
with a criterion for the applicability of the double scaling limit;
the $m$-th critical continuum theory exists iff dimensionality of
the target space satisfies $D<2(m+1)/(m-1).$

So far we have constrained ourselves to the {\it canonical}
scaling limit
of $O(N)$ vector field theories. This does not suffice to
render the contribution of an arbitrarily looped configuration of polymers
finite, because of ultraviolet divergences in the target space
with dimensionality $D \geq 2$.
In fact, the cosmological constant $t$ in the effective action (\ref{6.1.8})
must not be a constant, but be dependent upon the cutoff $\Lambda \sim a^{-1}$
in such a way that the counter-term part of $S_{eff} \BL \vp \BR$ cancels
divergences
originating from its physical part.
This means that the canonical double scaling limit (\ref{6.1.9})
where the combination
$\B\D^{\bigl( 2(m+1) - (m-1) D \bigr) /2m}$ is
kept fixed to
$\k^{-2} t^{\bigl( 2(m+1) - (m-1) D \bigr) /2m}$
is violated through the renormalization procedure \cite{ZJ2}.
In the following we clarify the situation by paying attention to
the {\it residual} divergences in the free energy
which is not explicitly stated in the above references.

As illuminating examples of the scaling violation,
we consider $\varphi^3$ field theories (\ref{6.1.6}) at $D=$1, 2, 3 and 4.
For convenience we employ the following form of the effective action
\beq
S_{eff} \left[ \vp \right] =
\int d^D \TX \lt
-\frac{1}{3} \frac{t^{3/2}}{\tk^2} + \frac{1}{2} (\tilde{\PA}_\mu \vp)^2
+ 2t^{1/2} \vp^2 + \frac{4\tk}{3} \vp^3 \rt
\label{6.2.1}
\eeq
by rescaling $\vp \too (2\pi)^{D/4} \vp$ ($\tk \equiv (2\pi)^{D/4} \k$).

For dimensions $D<6$, possible primitively divergent graphs are
zero-, one- and (the momentum-independent part of) two-point functions.
Thus we should separate the bare action (\ref{6.2.1}) into
the physical and the counter-term part appropriately as
\bea
S_{eff} &=&S_{phys} + S_{c.t.}\nonumber \\
&=&\int d^D \TX \lt
\frac{1}{2} (\TPA_\mu \vp_r)^2
+ 2t_r^{1/2} \vp_r^2 + \frac{4\tk}{3} \vp_r^3 \rt +
\int d^D \TX \lt
-c_0 - c_1 \vp_r -\HF c_2 \vp^2_r \rt .
\label{6.2.3}
\eea
Here $\vp_r$ is the renormalized field and $t_r$ is
the renormalized mass parameter\footnote{
Here the term `renormalized' is meant in a sense of target space,
but not of branched polymers.}.

What is important is that we are {\it not} necessarily allowed to
fix the renormalization coefficients $c_i (a)$
so that all divergences in {\it bubble diagrams} are canceled;
it is because the constant and the mass term
in the bare action (\ref{6.2.1}), to be equated to eq.(\ref{6.2.3})
through a shift in $\vp$ by a divergent constant, are related
as they have been extracted from the original vector model.
Since finiteness of correlation functions always takes preference over
that of vacuum bubbles in the renormalization procedure,
the free energy generically contains residual divergences.\\

{\sc One Dimension}\\
In one dimension the continuum theory is convergent so that the canonical
double scaling where $\B \D^{5/4}$ is kept fixed is maintained.
The free energy par unit volume reads
\bea
F &=& \sum_{g=0}^{\infty} \tk^{2g-2} \ t^{\frac{3}{2} - \frac{5}{4}g} \ f_g
\nonumber \\
  &=& -\frac{1}{3} \tk^{-2} t^{3/2} + t^{1/4}
  - \frac{11}{72} \tk^2 t^{-1} - \cdots
\eea
and the susceptibility exponents for the $g$-looped polymers are equal to
$\gamma^{(g)} = \HF + \frac{5}{4}g$.

\

{\sc Two Dimensions}\\
In two dimensions the one-loop tadpole is the only primitively divergent
correlation function, and among vacuum bubbles the one-loop bubble is
divergent.
The divergent tadpole is canceled by the counterterm
\beq
c_1(a) = -\frac{\tk}{2\pi} \ln (16 t_r a^4)
\eeq
and $c_2 =0$.
Equating eq.(\ref{6.2.3}) with eq.(\ref{6.2.1}), we find that the renormalized
quantities are related to the bare ones by
\beq
\vp = \vp_r + \frac{t_r^{1/2} - t^{1/2}}{2\tk}, \ \ \
t = t_r -\frac{\tk^2}{2\pi} \ln (16 t_r a^4) .
\label{6.2.5}
\eeq
After renormalization the free energy is completely finite
for each loop and is given by
\bea
F &=& \sum_{g=0}^{\infty} \tk^{2g-2}\  t_r^{\frac{3}{2} -g}\  f_g \nonumber\\
  &=& -\frac{1}{3} \tk^{-2} t^{3/2}_r + \frac{1}{2\pi} t^{1/2}_r
  - \frac{\alpha}{6\pi^2} \tk^2 t^{-1/2}_r - \cdots
\eea
where $\alpha \equiv \int^1_0 du (-\ln u)/(u^2-u +1) = 1.1719\cdots$.
The susceptibility exponents for the $g$-looped polymers, defined
with respect to $t_r$, are equal to $\gamma^{(g)} = \HF +g$.

If we recover the bare variables (in a sense of branched polymers)
$\B$ and $\D$ by eq.(\ref{6.1.5}), the second equation in eq.(\ref{6.2.5})
reads
\beq
\B\D= \frac{1}{4\pi} \ln \B +
\frac{t_r}{2\tk^2}-\frac{1}{4\pi}\ln \frac{t_r}{2\tk^2} .
\label{6.2.7}
\eeq
Thus we observe that the canonical scaling $\B\D = {\rm fixed}$ is violated,
instead we have to tune $\B \D /\ln \B$ to $1/4\pi$ in order to
render the continuum theory finite.
We remark that only the critical value of the combination
$\B \D /\ln \B$ (or equivalently $\B\D/\vert \ln \D \vert$)
is universal, while the way how its deviation from
$1/4\pi$ is related to the renormalized parameter $t_r$
depends on the regularization and renormalization schemes employed.

The logarithmic violation of the canonical combination of
scaling parameters is also observed in the non-perturbative
theory of strings in one dimension \cite{GMj,Pol}.
It is instructive to point out the similarities between
the two cases, that
scaling violations are induced by the tadpole of massless modes ---
$\vp$-field in our case and the tachyon field in the string theory
which becomes massless in one dimension --- in
effectively two-dimensional theories;
both represent the fluctuations of ($O(N)$- and $U(N)$-)invariant
composite fields.

\

{\sc Three Dimensions}

In three dimensions the one-loop tadpole is again the
only primitively divergent correlation function, and
the two-loop bubble
in addition to
the one-loop is also divergent.
The divergent tadpole is canceled by the counterterm
\beq
c_1 (a)=\frac{2\tk}{\pi^2} (a^{-1} -\pi t_r^{1/4})
\eeq
and $c_2 = 0$. We find that, by renormalizing $t$ as
\beq
t=t_r +\frac{2\tk^2}{\pi^2} (a^{-1} -\pi t_r^{1/4}) ,
\label{6.2.9}
\eeq
the free energy takes the almost finite form
\bea
F &=& \sum_{g=0}^{\infty} \tk^{2g-2}\  t_r^{\frac{3}{2} -\frac{3}{4}g}\  f_g
\nonumber \\
  &=& -\frac{1}{3} \tk^{-2} t_r^{3/2} + \frac{1}{3\pi} t_r^{3/4}
 +\frac{1}{6\pi^2} \tk^{2} \ln (16 t_r a^4)
  - \cdots
\eea
with one logarithmically divergent term ($F_2$), originating from the
above mentioned graph.
The susceptibility exponents for the $g$-looped polymers
are equal to $\gamma^{(g)} = \HF +\frac{3}{4}g$.

If we recover the bare variables, eq.(\ref{6.2.9})
reads
\beq
\B\D= \frac{1}{\pi^2} +
\B^{-1/3} \lt \frac{t_r}{(2\tk)^{8/3}}
-\frac{1}{2\pi}\frac{t_r^{1/4}}{(2\tk)^{2/3}} \rt .
\label{6.2.11}
\eeq
Thus we observe that the canonical scaling $\B\D^{3/4} = {\rm fixed}$
is violated with a power in $\B$. We have not yet
encountered a counterpart of the powerlike violation of scaling in the
non-perturbative theory of strings.

\

{\sc Four Dimensions}

In four dimensions the one-loop tadpole and the self energy are primitively
divergent correlation functions, and one-, two- and three-loop
bubbles contain divergences.
The divergence in correlation functions are canceled by the counterterms
\beq
c_1 (a)
=\frac{\tk}{4\pi^2} \lt a^{-2} +2 t_r^{1/2} \ln \lt 16 t_r a^4 \rt \rt,
\ \ \
c_2 (a)
=
\frac{\tk^2}{\pi^2} \ln \lt 16 t_r a^4 \rt .
\eeq
We find that, by renormalizing $t$ as
\beq
t=t_r +\frac{\tk^2}{4\pi} a^{-2} + \frac{\tk^4}{16\pi^2}
\ln^2 (16 t_r a^4) ,
\label{6.2.13}
\eeq
the free energy takes the almost finite form
\bea
F &=& \sum_{g=0}^{\infty} \tk^{2g-2} \ t_r^{\frac{3}{2} -\frac{1}{2}g}\  f_g
\nonumber\\
  &=& -\frac{1}{3} \tk^{-2} t_r^{3/2} + \frac{t_r}{8\pi^2} \lt \frac{1}{3}
\ln (16 t_r a^4) -1 \rt
  - \cdots
\eea
with three divergent terms (logarithmically divergent $F_1$,
quadratically divergent $F_2$ and logarithmically divergent $F_3$),
originating from the above mentioned graphs but diverging more mildly
than the unrenormalized ones.
The susceptibility exponents for the $g$-looped polymers
are equal to $\gamma^{(g)} = \HF +\frac{1}{2}g$.

If we recover the bare variables, eq.(\ref{6.2.13})
reads
\beq
\B\D= \frac{1}{8\pi^2} +
\B^{-1} \lt
\frac{t_r}{64\tk^4}+\frac{1}{1024\pi^4} \ln^2 \lt
\frac{t_r}{64\tk^2} \B^{-2} \rt \rt .
\label{6.2.15}
\eeq
Thus we observe that the canonical scaling $\B\D^{1/2} = {\rm fixed}$
is violated with a power in $\B$.

Up to now we have confirmed the following
expression of the susceptibility
exponent (defined with respect to $t_r$)
\beq
\gamma^{(g)}_{m,D} = 1-\frac{1}{m} +
\lt 1+\frac{1}{m} -\frac{m-1}{2m}D \rt g ,
\eeq
advocated in ref.\ \cite{ADJ} holds for all theories in concern.
According to the above formula, the exponent $2-\gamma$ happens to be a
nonnegative integers for several low-loop levels
($\gamma^{(2)}_{2,D=3}=\gamma^{(3)}_{2,D=4}=2$ and
$\gamma^{(1)}_{2,D=4}
=1$).
In these cases we have observed logarithmic
residual divergences in the free energies
as it should be, otherwise they could not depend nontrivially upon
the renormalized cosmological constant $t_r$.
We note that this situation is shared by the strings in one dimension,
where sphere and torus free energies (with exponents
$\gamma=0$ and 2, respectively) are logarithmically
divergent in the double scaling limit.\\

\indent
The author gratefully acknowledges T.\ Yoneya, M.\ Kato and N.\ Sakai
for helpful discussions and comments.

\baselineskip=0.6cm
\newcommand{\NP}{{\sl Nucl. Phys.\ }}
\newcommand{\PL}{{\sl Phys. Lett.\ }}
\newcommand{\MPL}{{\sl Mod. Phys. Lett.\ }}
\newcommand{\IJMP}{{\sl Int. Journ. Mod. Phys.\ }}
\newcommand{\IJ}{{\sl Int. Journ. Mod. Phys.\ }}
\newcommand{\CMP}{{\sl Commun. Math. Phys.\ }}
\newcommand{\PTP}{{\sl Prog. Theor. Phys.\ }}
\newcommand{\PTPS}{{\sl Prog. Theor. Phys. Suppl.\ }}
\newcommand{\PR}{{\sl Phys. Rev.\ }}
\newcommand{\PRL}{{\sl Phys. Rev. Lett.\ }}
\newcommand{\UMN}{{\sl Usp. Matem. Nauk.\ }}
\newcommand{\JP}{{\sl Journ. Phys.\ }}
\newcommand{\NC}{{\sl Nuovo Cimento\ }}

\end{document}